\documentclass[reprint]{revtex4-1}
\usepackage{graphicx}
\usepackage{dcolumn}
\usepackage{bm}
\usepackage{amsmath}
\usepackage{amssymb}

\begin{document}

\preprint{}

\title{Secure Text Mail Encryption with Generative Adversarial Networks}

\author{Dr.~Alexej~Schelle}
\affiliation{Constructor University, Bremen gGmbH, Campus Ring 1, 28759 Bremen, Germany}
\affiliation{IU Internationale Hochschule, Juri-Gagarin-Ring 152, D-99084 Erfurt, Germany}

\date{\today}

\begin{abstract}

This work presents an encryption model based on Generative Adversarial Networks (GANs). 
Encryption of RTF-8 data is realized by dynamically generating decimal numbers that lead to the encryption and decryption of alphabetic strings 
in integer representation by simple addition rules, the modulus of the dimension of the considered alphabet.
The binary numbers for the private dynamic keys correspond to the binary numbers of public reference keys, as defined by a specific GAN configuration.   
For reversible encryption with a bijective mapping between dynamic and reference keys, as defined by the GAN encryptor, secure text encryption can be achieved by transferring a GAN-encrypted public key along with the encrypted text from a sender to a receiver.
Using the technique described above, secure text mail transfer can be realized through component-wise encryption and decryption of text mail strings, with total key sizes of up to $10^{8}$ bits that define random decimal numbers generated by the GAN.
From the present model, we assert that encrypted texts can be transmitted more efficiently and securely than from RSA encryption, as long as users of the specific configuration of the GAN encryption model are unaware of the GAN encryptor circuit and configuration, respectively. 

\begin{description}
\item[Purpose] Whitepaper on arXiv.org (2025); Preprint for publication.
\end{description}
\end{abstract}

\maketitle

\section{Introduction}

Modern hybrid models for secure text mail encryption are based on the connection of symmetric and asymmetric encryption techniques using algorithms such as the Rivest-Shamir-Adleman encryption algorithm (RSA) to send encrypted keys from a sender to a receiver \cite{rivest1978method}.
Those methods usually rely on the definition of one public and one private key for secure key exchange, which a network may use to encrypt and decrypt a certain alphanumeric text mail structure \cite{boneh2003identity}.
Hybrid models do have the advantage of secure key transfer at the cost of lower computational efficiency; however, the generation of the key may be unsecured due to a lack of internal security.  
Additionally, the components of the secured key are encrypted with only the same single key.

To examine hybrid encryption models and their vulnerabilities, including the efficiency trade-offs between symmetric and asymmetric encryption techniques, the concept of Generative Adversarial Networks (GAN) proves to be a well-suited technique to model data that approximates the original data or data objects from the generation and comparison of the data objects 
modeled by the GAN with the original object \cite{goodfellow2014generative}.
A large range of applications has been developed with GANs in information sciences, such as sound or color recognition \cite{donahue2019adversarial, zhang2016colorful}, or more complex classification algorithms that 
serve to classify data structures and, in general, the elements of information theory \cite{karras2019style, zhu2017unpaired, brock2018large}.

Very fundamentally, Generative Adversarial Networks build an important component to understanding the functionality of biological and physical processes 
of neuronal networks.
In the framework of basic field theories for the modeling of information processing, the very basic mechanisms of learning as described by simple mathematical 
models such as the single-layer perceptron (SLP) work equivalently to the stepwise convergence of a GAN that finds configurations of information compatible 
with already known or consistent results \cite{rosenblatt1958perceptron, minsky1969perceptrons}. 
While technologies based on GANs have been developed quite extensively for applications in the context of artificial intelligence, 
fewer efforts have been made to enhance information security and data encryption techniques \cite{hu2022generative}.

As a natural type of use, GANs can be applied to find random configurations of binary numbers that vary with the internal modeling parameters of the artificial network.
In standard types of applications, the principle structure of a GAN is composed of a generator that models certain configurations of data structures and objects, respectively, 
which is connected to a differentiator that compares the generated data structures to pre-defined reference objects.
In this configuration, a GAN primarily builds an artificial learning method that belongs to the class of supervised learning algorithms. 

In the present work, we show how to avoid these weak points with key encryption models based on GANs that do not rely on factorization on the one hand.
Developing an encrypting technique that defines different components (decimal numbers) for the encryption of the key itself with a circuit encryptor configuration that is hidden from the users of the secure key additionally enhances the security of the encryption technique.  
Applying this GAN-based secure text mail encryption algorithm, we find that up to 24-bit encryption corresponding to total key sizes of up to $10^{5} - 10^{6}$ bits for a standard text mail of a few hundred words can be processed on a standard personal computer on the scale of a few minutes of computation time.
The security of text mail encryption with GANs is not obtained from the complexity of the key or the encryption algorithm alone but from the possibility of disconnecting (securing) the GAN encryptor from the parties that use the encryption model and prohibiting the access of external users to the actual specific random configuration of the (pre-defined) GAN encryptor. 

Encryption of RTF-8 (text) data is realized by dynamically generating private decimal numbers that lead to the encryption and decryption of alphabetic strings 
in integer representation by simple addition rules, the modulus of the dimension of the considered alphabet \cite{katz1996handbook, stinson2005cryptography}.
Random decimal values for the encryption of alphabetic components for email strings are calculated by the randomization of binary numbers with definite dimensions $N$ that generate decimal values with a random number generator in Python.  
The binary numbers for the private dynamical keys correlate with the binary numbers of public reference keys from a mapping defined by the specific GAN implementation.   
For reversible encryption with bijective mapping between dynamic and reference keys as defined by the GAN, the encoded text mail is transferred from a sender to a receiver together with the reference keys that build the basis for modeling decimal numbers from configurations of random binary keys.
Using the technique described, secure text mail transfer can be realized by component-wise encrypting text mail strings with random decimal numbers obtained from the GAN.

\begin{figure}[t]
    \centering
    \includegraphics[width=0.45\textwidth]{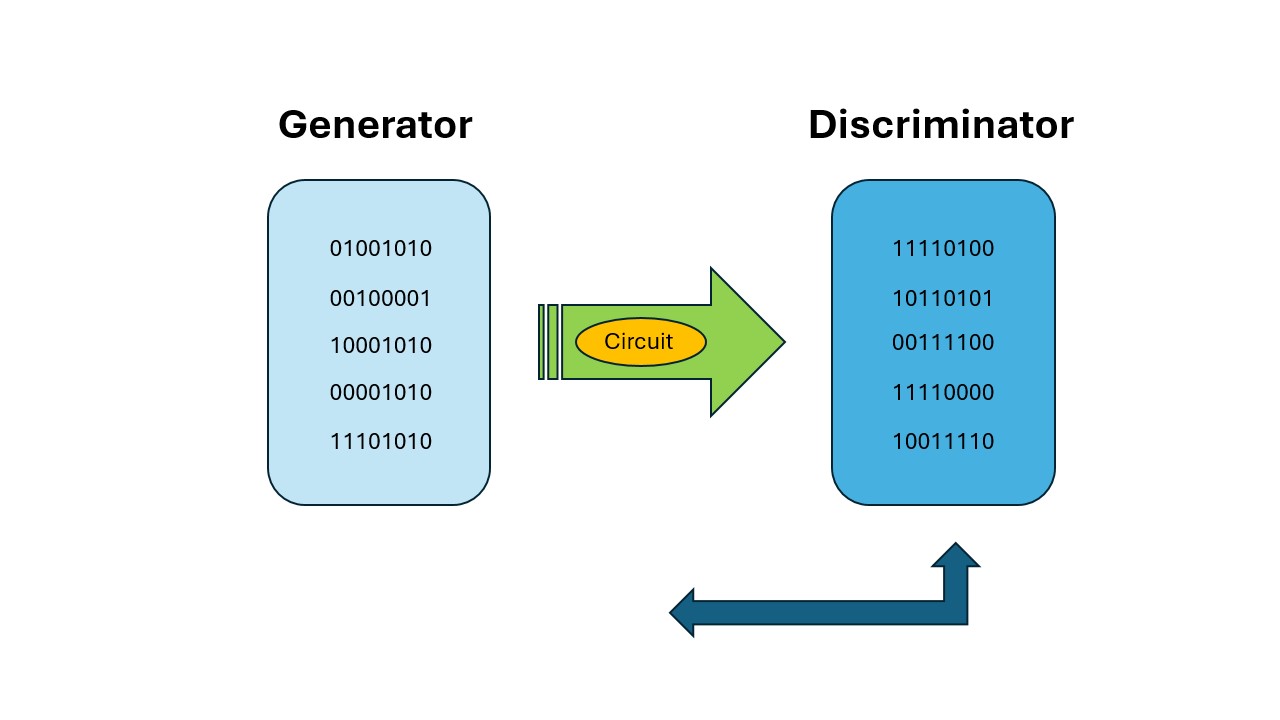} 
    \caption{(Color online) The conceptual setup of the Generative Adversarial Network (GAN) is shown in the figure above. The generating part of the technology builds dynamic N-bit-sized key values $G$ that are processed with the circuit of the GAN. From that configuration, reference keys $R$ are obtained at the discriminator of the GAN. The transmitted reference keys are secure against hacking since the circuit technology is generated randomly and, in particular, hidden from external observers. After receiving the (dynamically) encrypted text message with the reference keys, transmitted text structures such as emails or passwords can be decrypted with the recalled dynamical keys obtained intrinsically from the submitted reference keys by applying the GAN circuit technology.}
     \label{figure_1}
\end{figure}

\section{Theory}

\begin{figure}[t]
    \centering
    \includegraphics[width=0.45\textwidth]{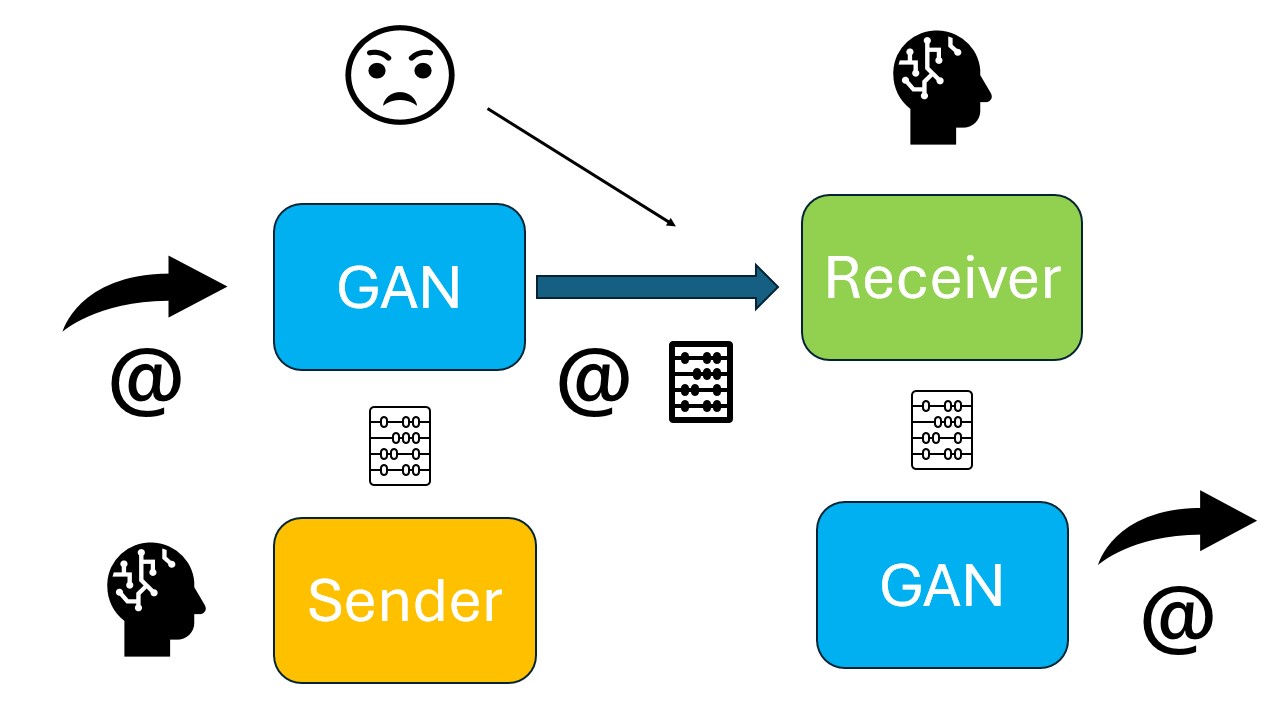} 
    \caption{(Color online) The GAN encryption model is illustrated in the figure above. 
A sender provides the text message of length $M$ that is sent from the sender to the GAN that encrypts the message with an (exemplary) randomly generated 
configuration of a total number of $L$ logical NOT gates. 
The logical NOT gates connect the N-bits of a dynamical key at the generator with the N-bits of the reference keys at the discriminator of the neuronal network that tests the total configuration for deviation until the checksum of the deviation equals zero. 
The encrypted message, together with the $M$ reference keys (that build a total key size of $M\times N$ bits) is sent to the receiver, which itself decrypts the message with $M$ dynamical keys obtained from applying the GAN to generate them from the reference keys for decryption. 
From simply pishing the encrypted text and the reference keys, an external observer isn't able to decrypt the message within polynomial time since the number of possible keys scales exponentially as $2^{M \times N}$.}
    \label{figure_2}
\end{figure}

Our encryption technology is built on two different components - the discriminator and the generator that exchange information in terms of N-bit binary number codes.
From the standard definition of GANs, we formally simplify our system parameters from setting all weighting coefficients of the two interacting (SLP) neuronal networks to the binary basis, i.e., $2^k$ \cite{ASHA2020}.
Each component of the GAN thus defines a decimal number from the N-bit signal, which in particular enables the definition of a complex number from the sum

\begin{equation}
m\left(G, R\right) = \sum^{N}_{j=1} a_j 2^j + i \sum^{N}_{j=1} b_j 2^j ,
\end{equation}\\
where $a_j, b_j\in\mathbb{Z}^N_2$ and $G, R \in \mathbb{Z}^N_2$. 
The mathematical function $m:\mathbb{Z}^N_2\times\mathbb{Z}^N_2\rightarrow\mathbb{C}$ formally maps two N-bit configurations of the keys $G = [a_1, ..., a_N] $ and $R = [b_1, ..., b_N]$ to the complex number space $\mathbb{C}$.
As shown in Fig. \ref{figure_1}, standard configurations of random numbers generated by the GAN generator can be modified with a circuit connected between the generator and the discriminator part of the encryption network.
Each realization of a randomly generated N-bit signal at the generating part of the GAN is processed in the circuit and then transferred to the differentiator, which distinguishes the generated signal from predefined reference signals or data patterns (see Fig. \ref{figure_2}).

Different circuit layers that are reversible concerning the mathematical transformation of the associated N-bit system can be configured for textual decryption.
An important formal aspect is the circuit's reversibility, which is expressed by bijective mathematical mappings that describe the interaction of the generator with the discriminator of the GAN.
Circuits can be built by connecting the standard logical NOT gate for reversible transformations and the logical gates AND, OR, NOR, and XOR, which leads to irreversible circuits for the generation of complex numbers (decimal number pairs) for the encryption and decryption of text structures \cite{Mano2015}.
Ensuring reversibility means that the GAN technology enables unique encryption and decryption of an associated RTF-8 text (compare Table I).

Similar to the RSA algorithm, we assume one public and one private key, where the public key is transferred from a sender to a receiver, and the private key applies to encrypt and decrypt the text message.
Instead of encrypting the text message or a transferred key with the (decrypted) combination of public and private keys (RSA), relying in particular on costly computation power, the security of the GAN approach makes use of the following technique.
Relevant text messages are connected to a GAN encrypter that entails a total number of $M$ reference keys (public keys) and $M$ dynamical keys (private keys) of size N-bit, i.e., total keys size of M$\times$N-bits, where $M$ is the number of letters in the text.
Each text message letter is encrypted and decrypted using a single character of the $ M$-bit dynamic N-bit keys for encryption and decryption, respectively.
Encryption and decryption are performed by mapping the relevant characters $A_k$ of the text message to a decimal number $B_k$ and adding the modulus of the decimal number for encryption and decryption, respectively, following the equation

\begin{equation}
B_k = (f(A_k) +{\rm Re}[m\left({\rm GAN}_c\left(G, R\right)\right)]~{\rm mod~} K 
\end{equation}\\
for encryption and 

\begin{equation}
A_k = (f^{-1}((B_k) - {\rm Re}[m\left({\rm GAN}_c\left(G, R\right)\right)]~{\rm mod ~} K
\end{equation}\\
for decryption, where ${\rm GAN}_c$ is a function that maps the N-bit configurations of the Generative Adversarial Network for a pair of N-bit configurations at the generator and discriminator, respectively, to a specific (relative) reference key $R^{\prime}$, i.e.

\begin{equation}
(G, R^{\prime}) = {\rm GAN}_c(G, R) ,
\end{equation}\\
given a certain configuration $C$ of the circuit technology, with $K$ the number of characters of the considered alphabet.
The function $f:\left\lbrace a, ..., Z\right\rbrace\rightarrow \mathbb{N}$ maps alphabetical values from the alphabet to numeric numbers $\mathbb{N}$ (decimal codes).
This way, each letter of the clear text is encrypted with a decimal number ${\rm Re}[m\left({\rm GAN}_c\left(G, R\right)\right)]\in\mathbb{N}$ obtained from the dynamic key, which is related to the reference key by the GAN that models each dynamical key $G \in \mathbb{Z}^N_2$ from a random reference key intrinsically resulting in a related random number ${\rm Im}[m\left({\rm GAN}_c\left(G, R\right)\right)]\in\mathbb{N}$. 

\renewcommand{\arraystretch}{1.2}

\begin{table}[t]
    \centering
    \begin{tabular}{|c|c|c|c|c|c|c|}
        \hline
        A, B & AND & OR & NOR & XOR & NOT\\
        \hline
        0, 0 & 0   & 0  & 1   & 0   & 1, 1 \\
        0, 1 & 0   & 1  & 0   & 1   & 1, 0 \\
        1, 0 & 0   & 1  & 0   & 1   & 0, 1 \\
        1, 1 & 1   & 1  & 0   & 0   & 0, 0 \\
        \hline
        C-time & 316.87  & 314.37  & 321.07  & 318.14 & 533.46 \\
        \hline
    \end{tabular}
    \caption{(Color online) Table summarizes the most important logical gates and the associated computation time for a GAN encryption model used to generate the (pairwise) circuit logic of the GAN for the encryption of a text message with a total of $3000$ characters for in total $M$ logical gates per GAN encryptor realization with $N$ bits, i.e. a total key size of $24\cdot10^{3}$ bits. Logical gates AND, NOR, OR, and XOR define irreversible logical operations. From the logical NOT gate, reversible text encryption is realized.}
\end{table}

The GAN encryption model enables secure text mail transfer by allowing the sender to access only the reference keys and the clear and encrypted text for submission to a receiver.
Encrypted text structures sent to the receiver are decrypted by connecting the reference keys with the GAN to generate the (same) dynamical key structure to decrypt the transferred text message.
Thereby, the encryptor configuration is unknown to the users of the GAN (sender and receiver), while the current configuration of the random circuit is (assumed to be) hidden from all external parties.
Compared to algorithms like RSA, besides a typically larger key size that is encrypted with the GAN encryptor, security is further enhanced by sending (a) decrypted reference key(s) on the one hand. 
The GAN setup itself allows the decryption of a text message only by decrypting the reference keys with the GAN encryptor that is hidden from the external observer.
Ideally, the security setup can be extended by securing the GAN encryptor from the sender and the receiver, e.g., building simple password protection for the Python source code, making the encryption model secure against internal information leaks.
Thus, an external observer cannot decrypt a text message by passing the password-protected key generation framework implemented in the Python programming language since the current circuit configuration is still unknown.
Passwords can be chosen to any complexity, and access to the system generating source code can be tracked, such as to change the encryptor configuration after access or after a certain period, with possible several attempts to access the key-generating Python framework \cite{VanRossum2009}.

We have tested the algorithm against scaling and find that $M$ times 8-bit to 24-bit key encryption and decryption corresponding to a total key size of up to $10^5 - 10^6$ digits for standard text messages of a few thousand ASCII signs are performed within a few seconds to minutes of computation time on a standard personal computer, depending on the key size of the component-wise GAN reference keys.
Figure 2 shows the scaling of the computational time against the number of bits used to generate keys for text encryption and decryption, respectively.  
The GAN algorithm works stably against failure modes and is available as a prototype software model on github.com \cite{TextmailEncryption}. 

\section{Results and Variations}

The main applicational scope of the present encryption algorithm is the secure text transfer from a sender to a receiver using the GAN encryptor as described in the previous chapter II.
Extensions for further applications like irreversible text deletion or password generation have been developed from the primary GAN encryption model, which has been tested to work successfully. 
Quantifying the computational performance of the routines in this applicational framework relies on testing different runtime variables as a function of the number of bits used to define the relevant decryption keys.  
For secure text mail encryption, scaling of computation time in terms of key complexity shows that a critical value of around a few hours of computation time is observed at component-wise key sizes of around 36 bits.
Instead of gaining security from large component-wise key sizes used for encryption, it is the size of the total key that defines a secure framework for information securing. 

Protecting information in the present security model further originates from protecting the randomly and automatically generated GAN encryptor from all parties involved in the text transfer process, i.e., a sender, 
a receiver, as well as a potential external hacker.
This is achieved by allowing access to the Python source code only for external third parties (to which current configurations of the randomly generated logical GAN circuit are unknown) rather than the actual users of the software and by automating the generation of the encryptor.
The structure of a GAN is built of a generator and a discriminator that builds an interacting system exchanging information as described above.
While the generating part tries to model N-bit-wise dynamical keys that fit a certain set of randomly created reference keys stored in the discriminator part of the GAN, a circuit 
that is configured between the two components ensures non-symmetric key structures between dynamic and reference keys.
Dynamic keys, which are pairwise connected to reference keys by the GAN encryptor, are used to encrypt and decrypt the relevant text in RTF-8 format.

In the present setup, we have defined logical operators mainly from combinations of NOT gates that ensure reversible transfer of N-bit logical signals as described by bijective functions defined on binary logical sets.
Convergence is achieved by generating dynamic keys that match the reference keys at the discriminator after passing the logical circuit exactly, relating a pair of dynamic and reference keys in each convergent step 
of a multiple of several subsequent GAN iteration cycles.
For a text string of a total of $M$ letters and signs, equally many reference keys are generated within a few seconds to minutes in an N-bit GAN generator with typically $N = 18 - 24$ serving for information transfer between the sender and the receiver of an encrypted message.
As realized in our case study, we have transferred the encrypted texts, such as passwords, by standard email from different senders to receivers, together with the reference keys for decryption over a wide distance of several thousand kilometers.
Text decryption was possible within seconds to minutes using the decryption of encrypted text messages using the decryption mode of the GAN for the reference keys.
An artificial memory intelligence has been built in the generator part of the GAN that was able to reduce the number of iteration steps for convergence but not the computation time to accelerate the approach of a zero-sum state (between the generator and the discriminator).  

\begin{figure}[t]
    \centering
    \includegraphics[width=0.45\textwidth]{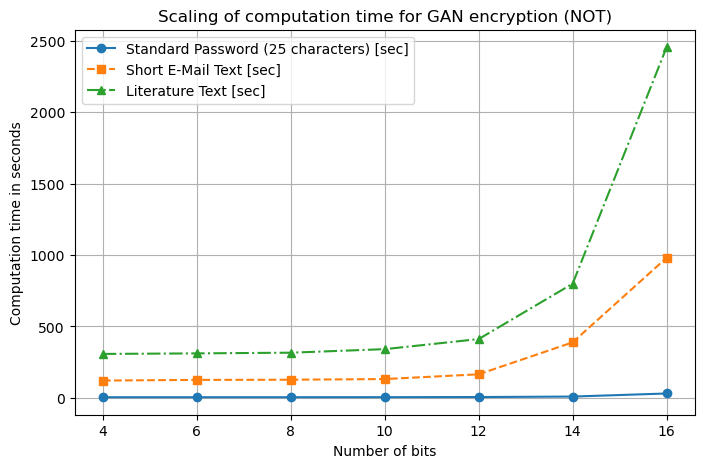} 
    \includegraphics[width=0.45\textwidth]{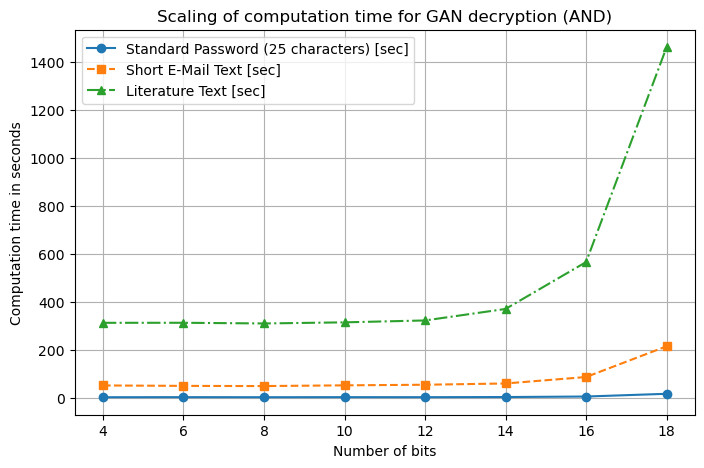} 
    \caption{(Color online) Shown is the scaling of computation time as a function of the number of bits (partial key sizes, i.e. one key per character of the encrypted message) used to \textit{reversibly} (upper figure) and \textit{irreversibly} (lower figure) encrypt standard passwords of length 25 characters (circles), small email text structures (squares) and one-page literature texts (triangles) - for connections realized by random configurations of NOT (upper figure) and AND (lower figure) logical gates integrated into the GAN encryptor.}
    \label{fig:figure_3}
\end{figure}

Mapping of characters to decimal number format from a function $f$ has been realized for text and numbers in RTF-8 format with dynamic keys of size N-bit (with values of $N$ up to $N=32$).
From an N-bit key structure, decimal numbers in the range of $0$ to $10^{8}$ are calculated by transforming binary key structures to decimal numbers.
Our source code has also been extended to recognize special characters in text strings or integer numbers and to distinguish letters case-sensitively.   
Counting the number of case-sensitive large and small letters is an important feature of the N-bit GAN encryption model.
In such a mode of operation, the GAN encryption model can be applied as a password validator and generator, respectively, that classifies passwords in terms of different complexity 
classes defined by the different variations of operation, i.e., class 1 for modes of special character recognition, class 2 for modes of number recognition, 
and class 3 for case-sensitive letter recognition. 
Most secure passwords can thus be generated, or detected, by randomizing character strings and combining randomly created strings that satisfy the conditions of classes 1 to 3.
Passwords of lower complexity are obtained from requiring the constraints for class 1 and/or class 2 conditions only.
Complexity 1 passwords of standard length (10 to 15 characters) have been generated and encrypted/decrypted within a few seconds of computation time, complexity 2 passwords within minutes, and finally, 
complexity 3 passwords within the time scale of less than an hour.

Finally, we have applied the GAN encryption model to perform non-reversible text deletion \cite{Shannon1948, Reardon2013}.
This mode of operation can be realized by implementing non-reversible circuit logic with the logical AND, OR, NOR, and XOR gates that are connected between the generator and the discriminator in the GAN encryptor.
Since the forward direction (encryption) of this mode of operation generates an ideally disjoint set of reference keys, 
encryption of a text string results in an irreversible mapping after the dynamic keys are overwritten by the encrypted keys - since, indeed in such case, the set of dynamic keys can not be remodeled from
the remaining set of (non-bijectively relating) reference keys in the backward direction (decryption).
 
\section{Discussion}

In the present approach, the concept of Generative Adversarial Networks applies to generate dynamic keys of large total size for encryption.
Text messages that can be read and mapped to numeric values from a standard text file are decrypted by calculating and modifying random decimal numbers with a random number generator that entails a circuit as an encryption technology.
The decrypted text message and the set of $M$ binary keys are sent from the sender to a receiver, enabling the decryption of the encrypted message with the receiver part.
The sender and the receiver, as much as an external hacker using the encryption software or phishing the encrypted text with large-size reference keys attached, respectively, are unable to access the Python script to which access is encrypted with a strong password, making the encryption model twofold secure against external hacking.   

Reversible encryption can be applied for email text exchange, secure password transfer or valid key generation.
Text mail structures in RTF-8 format of up to a few thousand words can be encrypted and securely forwarded using the GAN encryption model within a few seconds to minutes of computation time.  
Special characters that are mapped from object types to numeric numbers can be implemented independently and individually.  
Password encryption of complex passwords containing up to 100 or more characters can be realized as an integrated part of more complex encryption models for banking and insurance environmental applications \cite{corebanking, insurancetech, fintechsecurity}.

For such purposes, there are different approaches to integrating Python in a Java runtime environment that is suitable for programming user software applications \cite{pythonbook, javabook, pythonjavacomparison}.
Jython is an implementation of Python that allows for the integration of the programming language Python in a Java framework.
While all Python routines are accessible in the Jython environment, it only supports Python up to version 2.
Process builders or Java Native Interface with CPython do provide another application to call Python scripts from Java that is compatible with all versions of Python.
The programming language Python can also be integrated into other languages such as C/C++, JavaScript, .NET, and Go. 

Compared to the RSA algorithm, security is enhanced by first providing private and public keys that scale linearly with the number of digits used for password representation.
Starting from password sizes of $20 - 25$ digits, one may thus overbid the security of an RSA approach with partial key sizes of around $N = 16$ bits.
Applying encryption with $N = 24$ bits at password sizes of a few hundred digits, one can securely transfer passwords with total key sizes of up to a few thousand bits with the proposed GAN encryptor.      
Different and alternative encrypting algorithms such as ElGamal Encryption \cite{elgamal1985}, Elliptic Curve Cryptography \cite{koblitz1994}, or Lattice-Based Cryptography \cite{regev2005}, do work based on other computation methods but approximately provide the same security measure as the RSA algorithm in terms of encryption and decryption key sizes.    
Secondly, security is conceptually enhanced since the circuit logic of GAN isn't known to the users of the network, whereas the actual randomly generated configuration of logical gates is hidden from the developer of the software.
That way, encrypted text can only be hacked if two parties (software user and developer) lack information security.

\section{Conclusion}

In the present study, we have presented an encryption model for text encryption and decryption, respectively, based on a GAN encryptor that allows for secure RTF-8 text encryption. 
From modeling random keys that are unaccessible for decryption by any party of the GAN users without using the specific (protected) GAN encryptor model realization, i.e., sender, receiver, and software coordinator, the model implements the submission of encrypted text messages with reference keys from a sender to a receiver that build the foundation for decryption within the framework of the presented GAN prototype software.
Encryption and decryption with total key sizes of up to $10^6 - 10^8$ allow for secure text transfer and irreversible deletion of text structures or passwords in the framework of private as well as commercial applications and communication.
The model may build the foundation to develop commercial software technologies or integrated plugins in Python format from the presented basis model implemented in the Python 3 programming language.

\section{acknowledgments}

We thank Adrian Dahl, Sven Engels, Fritz Fischer, Mert Köktürk, Renars Miculis, Sarah Rosa Werner, and Betül Yurtman for their contributions and discussion on the content of the presented work on a text encryption model with Generative Adversarial Networks.

\bibliography{references}     

\end{document}